\documentclass[12pt]{article}
\begin{document}
\begin{center}

{\bf Pair Production and Vacuum Polarization of Vector Particles
with Electric Dipole Moments and Anomalous Magnetic Moments}\\
\vspace{5mm} S. I. Kruglov\\
\vspace{3mm}
\textit{International Educational Centre, 2727 Steeles Ave.West, Suite 202,\\
Toronto, Ontario, Canada M3J 3G9}\\
\vspace{5mm}
\end{center}

\begin{abstract}
The matrix, $8$-component Dirac-like form of $P$-odd equations for boson
fields of spins $1$ and $0$ are obtained and the $GL(2,c)$ symmetry group of
equations is derived. We found exact solutions of the field equation for
vector particles with arbitrary electric and magnetic moments in external
constant and uniform electromagnetic fields. The differential probability of
pair production of vector particles with the EDM and AMM by an external
constant and uniform electromagnetic field has been found using the exact
solutions. We have calculated the imaginary and real parts of the
electromagnetic field Lagrangian that takes into account the vacuum
polarization of vector particles.

\end{abstract}

\section{Introduction}

Vector particles W$^{\pm }$, Z$^0$ play very important role as
carriers of the weak interactions. The standard model of
electroweak interactions (SM) which implies the Higgs mechanism of
acquiring mass of vector particles is renormalizable.
Renormalizable electrodynamics for massive charged vector bosons
is based in the framework of the SM on the spontaneous breaking of
the local $SU(2)_L\otimes U(1)$ - symmetry. The $U(1)$ subgroup is
unbroken and the corresponding gauge electromagnetic field remains
massless. At the same time the gauge fields which are identified
with vector intermediate bosons (W$^{\pm }$, Z$^0$) corresponding
to broken $SU(2)_L$ - subgroup acquire masses. There is a certain
symmetry of the vector electromagnetic vertices in the
renormalizable SM and as a result the gyromagnetic ratio for
vector particles is equal to two. It should be noticed that for
the nonrenormalizable case of the Proca Lagrangian the
gyromagnetic ratio $g=1$. So, if an anomalous magnetic moment
(AMM) of vector particles is observed, which corresponds to $g\neq
2$, that would signal physics beyond the SM.

The $CP$ violation observed in the decays of the $K^0$ mesons and in $B_d^0/
\overline{B}_d^0\rightarrow J/\psi K_s^0$ decays remains mysterious. In the
SM $CP$ - violating interactions can be explained by the Kobayashi-Maskawa
mechanism which supposes single phase for three quark generations. In this
scheme the predicted electric dipole moments (EDM's) of elementary particles
are extremely small. In some supersymmetric and multi-Higgs models which are
extensions of the SM, $CP$ - violating effects are much stronger [1]. The
EDM of particles violates the time-reversal (T) symmetry and the $CP$
invariance which are equivalent due to the $CPT$ - theorem [2]. Some aspects
of the $CP$ - violating effects which follow from the EDM of the neutron,
electron and atoms are discussed in [3]. The EDM bounds of the neutron and
the electron can be established in low energy experiments.

There are some investigations of the EDM of vector W bosons in the
framework of the SM and beyond in [4]. But for W bosons it is
necessary to analyze the high energy processes for extracting $CP$
- odd asymmetries. The EDM of W bosons can give the large
contribution to the EDM of fermions (in particular to electrons).
The EDM of particles may be also induced by Higgs-boson exchange
[5]. The prediction of the EDM of the W-boson in the SM is
$d_W\simeq 10^{-29}$ $e$ cm [6] but beyond the SM it can be (for
example in the Two-Higgs-doublet model) about $10^{-21}-10^{-20}$
$e$ cm (see last reference in [4]). The experimental constraint on
the EDM of the W-boson which follows from the experimental upper
bound for the neutron EDM $ d_n=(-3\pm 5)\times 10^{-26}$ $e$ cm
[7] is $d_W\leq 10^{-19}$ $e$ cm. So, the presence of the EDM can
indicate physics beyond the SM.

Strong interacting composite hadrons $\rho $, $\omega $ (and
others) possess spin one. The theory of strong interactions of
quarks and gluons - quantum chromodynamics (QCD) is
renormalizable. However properties of hadrons are described by the
infrared region of QCD where perturbation theory in small
parameter $\alpha _s$ is not acceptable. In this region some
phenomenological models are used. The nonperturbative theory of
strong interactions of hadrons has not being developed yet but
there is a progress in describing hadrons in the framework of QCD
string theory [8]. In this approach the EDM of mesons [9] and
baryons [10] appears naturally. It should be noticed that the EDM
of the neutron may be induced by the $\vartheta $ -term of the QCD
vacuum. The QCD vacuum angle $\vartheta $ violates $P$ and $ CP$
symmetries and gives $CP$-odd electromagnetic observables. As the
EDM of the neutron is small the $\vartheta $-parameter of the QCD
vacuum is also small. It is possible to explore the axion
mechanism [11] to solve the strong $CP$ problem for having the
parameter $\vartheta =0$. Vector mesons may possess the EDM due to
the $\vartheta $-term [12] and the $CP$-odd electromagnetic
form-factors of $\rho $-mesons can be introduced.

In view of the great interest to physics in framework and beyond
the SM it is very important to study various processes involving
massive vector bosons with the EDM and AMM. The important and
interesting vacuum quantum effects are pair production of
particles and antiparticles and the vacuum polarization [13]. In
particular, there is the vacuum instability of the vector
particles in a magnetic field [14]. This is due to large
contribution of the tachyon mode to the negative part of the
Callan-Symanzik $\beta $-function, and as a result the vacuum is
reconstructed in a magnetic field. Some studies were performed to
investigate the vacuum quantum effects for vector fields. The pair
production and the vacuum polarization of vector fields with
gyromagnetic ratio $g=2$ by a constant uniform electric field were
investigated in [15]. The semiclassical imaginary-time method was
used in [16] to find the probability of pair production by a
constant electromagnetic field for arbitrary spin $s$ and
gyromagnetic ratio $g$. In [17] we found the pair production
probability and the vacuum polarization of fields for arbitrary
$s$ and $g$ on the basis of the exact solutions of the wave
equation  for particles in a constant and uniform electromagnetic
field and with the help of the Fock-Schwinger proper-time method.
In this approach fields realize $ (s,0)\oplus
(0,s)$-representation of the Lorentz group. Nonlinear corrections
to the constant uniform electromagnetic field due to the vacuum
polarization of a charged vector field in the framework of the
renormalizable gauge theory were studied in [18]. The pair
production probability of charged vector bosons with $g=1$ by a
non-stationary electric field was derived in [19].

In this work we study the pair production probability and the
vacuum polarization of the charged vector particles with arbitrary
EDM and AMM. The paper is organized as follows. In Sec. II we
proceed from the Dirac-K\"ahler equations for boson fields of
spins one and zero. We show that this system of wave equations can
be represented as two subsystems of $P$-odd equations for
self-dual and antiself-dual antisymmetric tensors of second rank.
The matrix form and the symmetry group of equations is
investigated in Sec. III. In Sec. IV the $P$-odd system of first
order equations for vector fields with the EDM and AMM is
introduced. We found exact solutions of the second order field
equation in external constant and uniform electromagnetic fields.
The pair production probability of vector particles with the EDM
and AMM is calculated in Sec. V with the help of found solutions.
Sec. VI devoted to finding the vacuum polarization of vector
particles. Section VII contains the conclusion.

\section{Field equations}

One of the nonperturbative approaches of strong interactions is
lattice QCD [20]. For describing fermions on the lattice the
Dirac-K\"ahler equation [21] can be used (see [22]).
Dirac-K\"ahler's equation in four dimensional space-time is given
by

\[
\left( d-\delta +m\right) \Phi =0,
\]

where $d$ is the exterior derivative, $\delta =-\star^{-1} d\star
$ transforms $ n- $form into $(n-1)-$form, $ \Phi $ denotes the
inhomogeneous differential form. The star operator $\star $
connects a $n-$form to a $(4-n)-$form so, that $\star ^2=1$,
$d^2=\delta ^2=0 $. The Laplacian is given by $ \left( d-\delta
\right) ^2=-\left( d\delta +\delta d\right) =\partial _\mu
\partial _\mu$ where the operator $\left( d-\delta \right)$
is the analog of the Dirac operator $\gamma _\mu \partial _\mu $.
The inhomogeneous differential form $ \Phi $ can be represented as
\[
\Phi =\varphi (x)+\varphi _\mu (x)dx^\mu +\frac 1{2!}\varphi _{\mu
\nu }(x)dx^\mu \wedge dx^\nu +
\]
\[
+\frac 1{3!}\varphi _{\mu \nu \rho }(x)dx^\mu \wedge dx^\nu \wedge
dx^\rho +\frac 1{4!}\varphi _{\mu \nu \rho \sigma }(x)dx^\mu
\wedge dx^\nu \wedge dx^\rho \wedge dx^\sigma,
\]

where $\wedge$ is the exterior product; $\varphi(x)$, $\varphi
_\mu (x)$, $\varphi _{\mu \nu }(x)$, $\varphi _{\mu \nu \rho
}(x)$, $\varphi _{\mu \nu \rho \sigma }(x)$ are scalar, vector and
antisymmetric tensor fields, respectively. The antisymmetric
tensors $ \varphi _{\mu \nu \rho }(x)$, $\varphi _{\mu \nu \rho
\sigma }(x)$ are connected with pseudovector and pseudoscalar
fields by the relationships:
\[
\widetilde{\varphi }_\mu (x)=\frac 1{3!}\varepsilon _{\mu \nu \rho
\sigma }\varphi _{\nu \rho \sigma
}(x),\hspace{0.5in}\widetilde{\varphi }(x)=\frac 1{4!}\varepsilon_
{\mu \nu \rho \sigma }\varphi _{\mu \nu \rho \sigma }(x),
\]
where $\varepsilon _{\mu \nu \alpha \beta }$ is an antisymmetric
tensor Levy-Civita; $\varepsilon _{1234}=-i$. The Dirac-K\"ahler
equation formulated in the framework of differential forms [21] is
equivalent to the following system of tensor fields [23]:
\begin{equation}
\partial _\nu \psi _{\mu \nu }(x)-\partial _\mu \psi (x)+m^2B_\mu (x)=0,
\hspace{0.5in}\partial _\nu \widetilde{\psi }_{\mu \nu }(x)-\partial _\mu
\widetilde{\psi }(x)+m^2C_\mu (x)=0,  \label{1}
\end{equation}
\begin{equation}
\partial _\mu B_\mu (x)-\psi (x)=0,\hspace{0.3in}\partial _\mu C_\mu (x)-
\widetilde{\psi }(x)=0,  \label{2}
\end{equation}
\begin{equation}
\psi _{\mu \nu }(x)=\partial _\mu B_\nu (x)-\partial _\nu B_\mu
(x)-\varepsilon _{\mu \nu \alpha \beta }\partial _\alpha C_\beta (x),
\label{3}
\end{equation}

where $\widetilde{\psi }_{\mu \nu }=(1/2)\varepsilon _{\mu \nu
\alpha \beta }\psi _{\alpha \beta }$ is the dual tensor.
Expression (3) is the most general representation for the
antisymmetric tensor of second rank [24, 25]. Equations (1)-(3)
describe the system of the vector ($B_\mu(x)$), pseudovector
($C_\mu(x)$), scalar ($\psi(x)$), and pseudoscalar
($\widetilde{\psi}(x)$) fields. For the complex values of the
vector-potentials $B_\mu(x)$ and $C_\mu(x)$, Eqs. (1)-(3)
correspond to the charged vector fields. As the system of
equations (1)-(3) contains two four-vectors $B_\mu(x)$, $C_\mu(x)$
which carry spins one and zero (without Lorenz conditions,
$\partial _\mu B_\mu (x)\neq 0$, $\partial _\mu C_\mu (x)\neq 0$)
there is the doubling of the spin states of particles. So. Eqs.
(1)-(3) describe fields with two spin-one and two spin-zero
states. Using the projection operator technique this states may be
separated [23]. The field equations (1)-(3) can be derived from
the corresponding Lagrangian and represent the Lagrange-Euler
equations. The Proca equations [26] are a special case of Eqs.
(1)-(3) when the constraints $C_\mu=0$, $\partial_\mu B_\mu =0$
are imposed. At the case of $C_\mu=0$, $\partial_\mu B_\mu \neq0$
we arrive at Stueckelberg's equation [27] describing spin one and
zero fields without the doubling of the spin states of a particle.
The matrix form of equations (1)-(3) is $ 16\times 16$
-dimensional Dirac equation [23]. This makes it possible to
describe fermions with spin $1/2$ with the help of fields $\psi(x)
$, $B_\mu(x)$, $\psi _{\mu \nu }(x)$, $\widetilde{\psi }(x)$,
$C_\mu(x)$ which do not realize the tensor representation of the
Lorentz group in this case and are connected with spinors. In this
work we consider the case when the fields $\psi(x)$, $B_\mu(x)$,
$\psi _{\mu \nu }(x)$, $\widetilde{\psi }(x)$, $C_\mu (x)$ are
bosonic fields carrying spins $0$ and $1$.

Dirac-K\"ahler equations (1)-(3) are equivalent to the following systems
\[
\partial _\nu M_{\mu \nu }(x)-\partial _\mu M(x)+m^2M_\mu (x)=0,
\hspace{0.3in}\partial _\mu M_\mu (x)=M(x),
\]
\vspace{-8mm}
\begin{equation}
\label{4}
\end{equation}
\vspace{-8mm}
\[
M_{\mu \nu }(x)=\partial _\mu M_\nu (x)-\partial _\nu M_\mu
(x)-i\varepsilon _{\mu \nu \alpha \beta }\partial _\alpha M_\beta
(x),
\]

with the self-dual tensor $M_{\mu \nu }(x)=-i\widetilde{M}_{\mu \nu }(x)$
and
\[
\partial _\nu N_{\mu \nu }(x)-\partial _\mu N(x)+m^2N_\mu (x)=0,
\hspace{0.3in}\partial _\mu N_\mu (x)=N(x),
\]
\vspace{-8mm}
\begin{equation}
\label{5}
\end{equation}
\vspace{-8mm}
\[
N_{\mu \nu }(x)=\partial _\mu N_\nu (x)-\partial _\nu N_\mu
(x)+i\varepsilon _{\mu \nu \alpha \beta }\partial _\alpha N_\beta
(x),
\]

with the antiself-dual tensor $N_{\mu \nu }(x)=i\widetilde{N}_{\mu
\nu }(x)$, where
\[
M(x)=\frac 1{\sqrt{2}}\left( \psi (x)-i\widetilde{\psi }(x)\right) ,
\hspace{0.3in}N(x)=\frac 1{\sqrt{2}}\left( \psi (x)+i\widetilde{\psi }
(x)\right) ,
\]
\begin{eqnarray*}
M_\mu (x) &=&\frac 1{\sqrt{2}}\left( B_\mu (x)-iC_\mu (x)\right) ,
\hspace{0.3in}M_{\mu \nu }(x)=\frac 1{\sqrt{2}}\left( \psi _{\mu \nu }(x)-i
\widetilde{\psi }_{\mu \nu }(x)\right) , \\
N_\mu (x) &=&\frac 1{\sqrt{2}}\left( B_\mu (x)+iC_\mu (x)\right) ,
\hspace{0.3in}N_{\mu \nu }(x)=\frac 1{\sqrt{2}}\left( \psi _{\mu \nu }(x)+i
\widetilde{\psi }_{\mu \nu }(x)\right) .
\end{eqnarray*}

Adding and subtracting equations (1)-(3) we get equations (4),
(5). The self-dual tensor $M_{\mu \nu }$ which obeys the equations
(4) is transformed under $\left( 1,0\right) $-representation of
the Lorentz group and has $3$ independent components (see also
[28]). Equations (4) are not invariant under the parity
transformation and there is no Lagrangian formulation of them.
This also applies to equations (5) for the antiself-dual tensor
$N_{\mu \nu }$ which transforms under the $\left( 0,1\right)
$-representation of the Lorentz group. But if we consider the
whole system of equations (4), (5) (which is equivalent to
equations (1)-(3)) on the basis of $(0,0)\oplus (1/2,1/2)\oplus
\left( 1,0\right) \oplus \left( 0,1\right) \oplus (1/2,1/2)\oplus
(0,0)$-representation of the Lorentz group, we will have
P-invariant theory within the Lagrangian formulation. Each of the
system of equations (4), (5) describes eight independent variables
($M(x)$, $M_\nu (x)$ , $M_{ab}(x)$), ($N(x)$, $N_\nu (x)$,
$N_{ab}(x)$).

\section{Matrix form of equations}

Let us introduce four-component columns:
\[
\xi (x)=-im\left(
\begin{array}{c}
M_a(x) \\
M_4(x)
\end{array}
\right) ,\hspace{0.3in}\chi (x)=\left(
\begin{array}{c}
\widetilde{M}_a(x) \\
M(x)
\end{array}
\right) ,
\]
\vspace{-8mm}
\begin{equation}
\label{6}
\end{equation}
\vspace{-8mm}
\[
\xi ^{\prime }(x)=-im\left(
\begin{array}{c}
N_a(x) \\
N_4(x)
\end{array}
\right) ,\hspace{0.3in}\chi ^{\prime }(x)=\left(
\begin{array}{c}
\widetilde{N}_a(x) \\
N(x)
\end{array}
\right) ,
\]

where $\widetilde{M}_a(x)=(1/2)\epsilon _{amn}M_{mn}(x)$, $\widetilde{N}
_a(x)=(1/2)\epsilon _{amn}N_{mn}(x)$. Taking into account the notations (6),
Eqs. (4), (5) can be represented as
\[
\alpha _\mu \partial _\mu \xi (x)=m\chi (x),
\]
\vspace{-8mm}
\begin{equation}
\label{7}
\end{equation}
\vspace{-8mm}
\[
\overline{\alpha }_\mu \partial _\mu \chi (x)=m\xi (x),
\]
\[
\alpha _\mu ^{\prime }\partial _\mu \xi ^{\prime }(x)=m\chi ^{\prime }(x),
\]
\vspace{-8mm}
\begin{equation}
\label{8}
\end{equation}
\vspace{-8mm}
\[
\overline{\alpha }_\mu ^{\prime }\partial _\mu \chi ^{\prime
}(x)=m\xi ^{\prime }(x),
\]

where
\[
\alpha _1=\left(
\begin{array}{cccc}
0 & 0 & 0 & -i \\
0 & 0 & -i & 0 \\
0 & i & 0 & 0 \\
i & 0 & 0 & 0
\end{array}
\right) ,\hspace{0.3in}\alpha _2=\left(
\begin{array}{cccc}
0 & 0 & i & 0 \\
0 & 0 & 0 & -i \\
-i & 0 & 0 & 0 \\
0 & i & 0 & 0
\end{array}
\right) ,
\]
\[
\alpha _3=\left(
\begin{array}{cccc}
0 & -i & 0 & 0 \\
i & 0 & 0 & 0 \\
0 & 0 & 0 & -i \\
0 & 0 & i & 0
\end{array}
\right) ,\hspace{0.3in}\alpha _1^{\prime }=\left(
\begin{array}{cccc}
0 & 0 & 0 & i \\
0 & 0 & -i & 0 \\
0 & i & 0 & 0 \\
i & 0 & 0 & 0
\end{array}
\right) ,
\]
\[
\alpha _2^{\prime }=\left(
\begin{array}{cccc}
0 & 0 & i & 0 \\
0 & 0 & 0 & i \\
-i & 0 & 0 & 0 \\
0 & i & 0 & 0
\end{array}
\right) ,\hspace{0.3in}\alpha _3^{\prime }=\left(
\begin{array}{cccc}
0 & -i & 0 & 0 \\
i & 0 & 0 & 0 \\
0 & 0 & 0 & i \\
0 & 0 & i & 0
\end{array}
\right) ,
\]
\vspace{-8mm}
\begin{equation}
\label{9}
\end{equation}
\vspace{-8mm}
\[
\alpha _4^{\prime }=\left(
\begin{array}{cccc}
-i & 0 & 0 & 0 \\
0 & -i & 0 & 0 \\
0 & 0 & -i & 0 \\
0 & 0 & 0 & i
\end{array}
\right) ,\hspace{0.3in}\overline{\alpha }_1^{\prime }=\left(
\begin{array}{cccc}
0 & 0 & 0 & -i \\
0 & 0 & -i & 0 \\
0 & i & 0 & 0 \\
-i & 0 & 0 & 0
\end{array}
\right) ,
\]
\[
\overline{\alpha }_2^{\prime }=\left(
\begin{array}{cccc}
0 & 0 & i & 0 \\
0 & 0 & 0 & -i \\
-i & 0 & 0 & 0 \\
0 & -i & 0 & 0
\end{array}
\right) ,\hspace{0.3in}\overline{\alpha }_3^{\prime }=\left(
\begin{array}{cccc}
0 & -i & 0 & 0 \\
i & 0 & 0 & 0 \\
0 & 0 & 0 & -i \\
0 & 0 & -i & 0
\end{array}
\right) ,
\]
\[
\overline{\alpha }_4^{\prime }=\left(
\begin{array}{cccc}
i & 0 & 0 & 0 \\
0 & i & 0 & 0 \\
0 & 0 & i & 0 \\
0 & 0 & 0 & -i
\end{array}
\right) ,\hspace{0.3in}\alpha
_4=iI_4,\hspace{0.3in}\overline{\alpha }_\mu =\left( \alpha
_k,-iI_4\right) .
\]

Equations (7), (8) can be also cast in the form
\begin{equation}
\beta _\mu \partial _\mu \varphi (x)+m\varphi (x)=0,  \label{10}
\end{equation}
\begin{equation}
\beta _\mu ^{\prime }\partial _\mu \varphi ^{\prime }(x)+m\varphi ^{\prime
}(x)=0,  \label{11}
\end{equation}

where
\[
\varphi (x)=\left(
\begin{array}{c}
\chi (x) \\
\xi (x)
\end{array}
\right) ,\hspace{0.3in}\beta _\mu =-\left(
\begin{array}{cc}
0 & \alpha _\mu \\
\overline{\alpha }_\mu & 0
\end{array}
\right) ,
\]
\vspace{-8mm}
\begin{equation}
\label{12}
\end{equation}
\vspace{-8mm}
\[
\varphi ^{\prime }(x)=\left(
\begin{array}{c}
\chi ^{\prime }(x) \\
\xi ^{\prime }(x)
\end{array}
\right) ,\hspace{0.3in}\beta _\mu ^{\prime }=-\left(
\begin{array}{cc}
0 & \alpha _\mu ^{\prime } \\
\overline{\alpha }_\mu ^{\prime } & 0
\end{array}
\right) ,
\]

and the matrices $\beta _\mu $, $\beta _\mu ^{\prime }$ obey the Dirac
algebra
\begin{equation}
\beta _\mu \beta _\nu +\beta _\nu \beta _\mu =2\delta _{\mu \nu }.
\label{13}
\end{equation}

We can combine Eqs. (10), (11) in the $16-$component Dirac-type wave
equation, as follows
\begin{equation}
\left( \Gamma _\mu \partial _\mu +m\right) \Psi (x)=0,  \label{14}
\end{equation}

where
\begin{equation}
\Psi (x)=\left(
\begin{array}{c}
\varphi (x) \\
\varphi ^{\prime }(x)
\end{array}
\right) ,\hspace{0.3in}\Gamma _\mu =\left(
\begin{array}{cc}
\beta _\mu & 0 \\
0 & \beta _\mu ^{\prime }
\end{array}
\right) .  \label{15}
\end{equation}

The $16\times 16$ - matrices $\Gamma _\mu $ also obey the Dirac algebra
(13). This means that the system of equations (7), (8) is equivalent to four
Dirac equations. So Dirac-K\"ahler equations are equivalent to two matrix
equations (10), (11) (or two systems of tensor equations (4), (5)). Equation
(10) (and (4)) as well as Eq. (11) (and (5)) are parity noninvariant
separately and at the same time the system of the two equations (10), (11)
(or Dirac-K\"ahler equations) are P-invariant.

Now we will find the symmetry group of equations (10), (11). As matrices $
\beta _\mu $, $\beta _\mu ^{\prime }$ obey the same algebra, equations (10)
and (11) have the same symmetry group. Therefore, we only need to consider
equation (10) which is equivalent to Eqs. (4).

The matrices $\beta _\mu $ are $8-$component Dirac-type matrices, and in a
specific basis they take the form $\widehat{\beta }_\mu =I_2\otimes \gamma
_\mu $. It is obvious that the matrices $\widehat{\overline{\beta }}_m=\tau
_m\otimes I_4$ ($\tau _m$ are the Pauli matrices) form the symmetry algebra
of Eq. (10). It should be noted that the internal symmetry under
consideration is not violated by introducing the electromagnetic fields by
the substitution $\partial _\mu \rightarrow \partial _\mu -ieA_\mu $. In the
representation (12), the following matrices
\begin{equation}
\overline{\beta }_m=\left(
\begin{array}{cc}
\rho _m & 0 \\
0 & \rho _m
\end{array}
\right) ,  \label{16}
\end{equation}

commute with the matrices $\beta _\mu $, if $\left[ \rho _m,\alpha _n\right]
=0$, where the matrices $\alpha _n$ are given by Eqs. (9) and satisfy the
Pauli commutation relations: $\left\{ \alpha _i,\alpha _k\right\} =2\delta
_{ik}$, $\left[ \alpha _i,\alpha _k\right] =2i\epsilon _{ikl}\alpha _l$.
Such matrices $\rho _m$ which commute with $\alpha _n$ have the form
\[
\rho _1=\left(
\begin{array}{cccc}
0 & 0 & 0 & -i \\
0 & 0 & i & 0 \\
0 & -i & 0 & 0 \\
i & 0 & 0 & 0
\end{array}
\right) ,\hspace{0.3in}\rho _2=\left(
\begin{array}{cccc}
0 & 0 & -i & 0 \\
0 & 0 & 0 & -i \\
i & 0 & 0 & 0 \\
0 & i & 0 & 0
\end{array}
\right) ,
\]
\vspace{-8mm}
\begin{equation}
\label{17}
\end{equation}
\vspace{-8mm}
\[
\rho _3=\left(
\begin{array}{cccc}
0 & i & 0 & 0 \\
-i & 0 & 0 & 0 \\
0 & 0 & 0 & -i \\
0 & 0 & i & 0
\end{array}
\right) ;
\]

they also obey the Pauli commutation relations. Further we will use also the
matrix $\beta _4=iI_4$.

Let us consider the group of the transformations of the wave function of Eq.
(10):
\begin{equation}
\varphi (x)\rightarrow \exp \left( m_\mu \overline{\beta }_\mu \right)
\varphi (x),  \label{18}
\end{equation}

where the $m_\mu $ are four complex parameters. The transformations (18) are
defined for the complex fields describing the charge fields; they form the
internal symmetry group of Eq. (10) which is isomorphic to the $GL(2,c)$
group.

It is possible to apply Eq. (10) to the description of spinor particles. In
this case the wave function $\varphi (x)$ realize the spinor representation
of the Lorentz group and Eq. (10) is equivalent to two Dirac equations; it
can be obtained by the variation procedure from the corresponding
Lagrangian. Thus generators of the Lorentz group are given by
\begin{equation}
J_{\mu \nu }^{(1/2)}=\frac 14\left( \beta _\mu \beta _\nu -\beta _\nu \beta
_\mu \right) ,  \label{19}
\end{equation}

and the Hermitianizing matrix is $\eta =\beta _4$.

In the case of the bosonic fields (see Eqs. (6), (12)), however, there is no
Lagrangian formulation of Eq. (10) because it is parity noninvariant
equation based on the reducible $(0,0)\oplus (1/2,1/2)\oplus $ $\left(
1,0\right) $-representation of the Lorentz group.

The requirement that the Lagrangian of spinor fields ($\eta =\beta _4$) be
invariant under the transformations (18) yields the restriction on the
parameters: $m_k^{*}=-m_k$, $m_4^{*}=m_4$, that corresponds to the
extraction of the $U(2)$ subgroup. According to the Noether theorem, this
produces the conservation current
\begin{equation}
\theta _{\mu \alpha }=\overline{\varphi }(x)\beta _\mu \overline{\beta }
_\alpha \varphi (x),  \label{20}
\end{equation}

so that $\partial _\mu \theta _{\mu \alpha }=0$;
$\overline{\varphi }(x)=\varphi^+(x)\beta_4$, $\varphi^+(x)$ is
the Hermitian-conjugate wave function. It is easy to verify that
the quantity (20) is also conserved in the boson case (see also
[28]), when the fields are given by Eqs. (6), (12). We notice that
the internal symmetry group of Dirac-K\"ahler equation (14) is
$GL(4,c)$ and the corresponding Lagrangian for bosonic fields is
invariant under the transformations of $ SO(4,2)$ group (or
locally isomorphic group $SU(2,2)$) [29, 23].

\section{Vector particle with EDM and AMM in uniform electromagnetic
field}

We consider here the description of electromagnetic interactions
of vector particles possessing the EDM and AMM. Sakata and
Taketani added some terms in equations which describe the effects
of anomalous moments [30], and Corben and Schwinger [31] included
the AMM in the Proca equations [26]. Yang and Bludman considered
an anomalous electric quadrupole moment [32]. Introducing in Eqs.
(4) the interaction with the electromagnetic field $
\partial _\mu \rightarrow {\cal D}_\mu =\partial _\mu -ieA_\mu $, AMM ,
and EDM, we arrive at the equations (at the substitutions $M\rightarrow \psi$,
$M_\mu \rightarrow \psi _\mu $, $M_{\mu \nu} \rightarrow \psi _{\mu \nu}$)
\[
{\cal D}_\mu \psi _\mu =\psi ,
\]
\begin{equation}
\psi _{\mu \nu }={\cal D}_\mu \psi _\nu -{\cal D}_\nu \psi _\mu
+\sigma \epsilon _{\mu \nu \alpha \beta }{\cal D}_\alpha \psi _\beta ,
\end{equation}
\[
{\cal D}_\nu \psi _{\mu \nu }-{\cal D}_\mu \psi +m^2\psi _\mu
+ie\kappa F_{\mu \nu }\psi _\nu =0.  \label{21}
\]

If setting $\sigma =0$ we get Stueckelberg's equations with the
AMM $e\kappa $ which describe fields of spin $1$ and $0$ [27]. It
should be noted that for field equations (21) the mass of the
field with spin of zero coincides with the mass of the vector
field. Quantization of the fields (21) leads to the indefinite
metric for scalar state. At $\sigma =i$ and $\kappa =0$ Eqs. (21)
have the $8-$component matrix formulation (10) (with the
replacement $
\partial _\mu \rightarrow {\cal D}_\mu $) with matrices $\beta _\mu $
(12) obeying the Dirac algebra. It is easy to obtain the second order
equation for the four-vector $\psi _\mu (x)$ from Eq. (21). As a result one
finds
\begin{equation}
\left( {\cal D}_\nu ^2-m^2\right) \psi _\mu (x)+ie\left( \sigma
\widetilde{F}_{\mu \nu }-gF_{\mu \nu }\right) \psi _\nu (x)=0,  \label{22}
\end{equation}

where $\widetilde{F}_{\mu \nu }=(1/2)\varepsilon _{\mu \nu \alpha
\beta }F_{\alpha \beta }$ is the dual tensor, $g=1+\kappa $ is the
gyromagnetic ratio for the quanta of spin $1$. Eq. (22) describes
a particle with the magnetic moment$\ eg/(2m)$ and the EDM $\sigma
/(2m)$. It should be noted that in the case of the Proca equation
with the EDM and AMM, we have in Eq. (22) the additional term
($-{\cal D}_\mu {\cal D}_\nu \psi _\nu $) due to the absence of a
scalar state. Eq. (22) can be treated in the framework of $\xi
$-formalism [33] as a wave equation for vector field in the gauge
$ \xi =1$. We notice that the formal counting of the divergences
corresponding to Eq. (22) leads to a renormalizable theory due to
the form of the field propagator which is proportional to $1/p^2$
but with the presence of indefinite metric.

It is easier to solve Eq. (22) compared to the Proca equation for a particle
in the external electromagnetic fields . To estimate the physical quantities
for a vector particle one needs to eliminate the contribution of a scalar
state. In the following calculations we will use this procedure.

Here we will find the solutions of Eq. (22) for a particle in the field of
uniform and constant electromagnetic fields. We note [13] that matrices $
F_{\mu \nu }$, $\widetilde{F}_{\mu \nu }$ have eigenvalues as follows
\[
F_{\mu \nu }\psi _\nu ^{(\lambda )}=F^{(\lambda )}\psi _\mu ^{(\lambda )},
\]
\[
\widetilde{F}_{\mu \nu }\psi _\nu ^{(\lambda )}=\frac 1{F^{(\lambda )}}
{\cal G}\psi _\mu ^{(\lambda )},
\]

\begin{equation}
F^{(\lambda )}=\pm F^{(1)},\pm F^{(2)},\hspace{0.3in}\lambda =1,2,3,4,
\end{equation}
\[
F^{(1)}=\frac i{\sqrt{2}}\left[ \left( {\cal F}+i{\cal G}\right)
^{1/2}+\left( {\cal F}-i{\cal G}\right) ^{1/2}\right] ,
\]
\[
F^{(2)}=\frac i{\sqrt{2}}\left[ \left( {\cal F}+i{\cal G}\right)
^{1/2}-\left( {\cal F}-i{\cal G}\right) ^{1/2}\right] ,  \label{23}
\]
\begin{equation}
{\cal F}=\frac 14F_{\mu \nu }^2=\frac 12\left( {\bf H}^2-{\bf E}
^2\right) ,\hspace{0.3in}{\cal G}=\frac 14F_{\mu \nu
}\widetilde{F}_{\mu \nu }={\bf E}\cdot {\bf H} ,  \label{24}
\end{equation}

and ${\bf E}$, ${\bf H}$ are the electric and magnetic fields,
respectively. In the diagonal representation (23) Eq. (22) becomes
\begin{equation}
\left( D_\nu ^2-m^2\right) \psi _\mu ^{(\lambda )}(x)+ie\left(
\sigma \frac 1{F^{(\lambda )}}{\cal G}-gF^{(\lambda )}\right) \psi
_\mu ^{(\lambda )}(x)=0 .  \label{25}
\end{equation}

Equation (25) represents the Klein-Gordon type equation for every component
of the eigenfunction $\psi _\mu ^{(\lambda )}(x)$. We consider the general
case when two Lorentz invariants of the electromagnetic fields ${\cal F}
\neq 0,$ ${\cal G}\neq 0$. It is convenient to use a coordinate system in
which the electric ${\bf E}$ and magnetic ${\bf H}$ fields are
parallel (${\bf E}={\bf n}E,$ ${\bf H}={\bf n}H$, ${\bf n}
=(0,0,1)$) and the 4-vector potential takes the form

\begin{equation}
A_\mu =\left( 0,x_1H,-tE,0\right) .  \label{26}
\end{equation}

After introducing the variables [34] (see also [35])
\[
\eta =\frac{p_2-eHx_1}{\sqrt{eH}},\hspace{0.5in}\tau
=\sqrt{eE}\left( t+ \frac{p_3}{eE}\right) ,
\]
\vspace{-8mm}
\begin{equation}
\label{27}
\end{equation}
\vspace{-8mm}
\[
\psi _\mu ^{(\lambda )}(x)=\exp \left[ i\left(
p_2x_2+p_3x_3\right) \right] \Phi _\mu ^{(\lambda )}(\eta ,\tau )
\]

Eq. (25) reads
\begin{equation}
\left[ eH\left( \partial _\eta ^2-\eta ^2\right) -eE\left( \partial _\tau
^2+\tau ^2\right) -m^2+ie\left( \sigma \frac 1{F^{(\lambda )}}{\cal G}
-gF^{(\lambda )}\right) \right] \Phi _\mu ^{(\lambda )}(\eta ,\tau )=0
\label{28}
\end{equation}

where $\partial _\eta =\partial /\partial \eta $, $\partial _\tau =\partial
/\partial \tau $. The solution to Eq. (28) exists in the form
\begin{equation}
\Phi _\mu ^{(\lambda )}(\eta ,\tau )=\xi _\mu \phi ^{(\lambda )}(\eta )\chi
^{(\lambda )}(\tau ),  \label{29}
\end{equation}

with a constant vector $\xi _\mu $, and the eigenfunctions $\phi
^{(\lambda )}(\eta )$, $\chi^{(\lambda )}(\tau )$ obey the
following equations
\begin{equation}
\left[ eH\left( \partial _\eta ^2-\eta ^2\right) -m^2+ie\left( \sigma \frac
1{F^{(\lambda )}}{\cal G}-gF^{(\lambda )}\right) +k_\lambda ^2\right]
\phi ^{(\lambda )}(\eta )=0,  \label{30}
\end{equation}
\begin{equation}
\left[ eE\left( \partial _\tau ^2+\tau ^2\right) +k_\lambda ^2\right] \chi
^{(\lambda )}(\tau )=0,  \label{31}
\end{equation}

where $k_\lambda ^2$ are the eigenvalues. The finite solution (at $\eta
\rightarrow \infty $) to Eq. (30) is
\begin{equation}
\phi ^{(\lambda )}(\eta )=N_0\exp \left( -\frac{\eta ^2}2\right)
H_n(\eta ), \label{32}
\end{equation}

where $N_0$ is the normalization constant, $H_n(\eta )$ are the
Hermite polynomials. The requirement that this solution be finite
leads to the condition
\begin{equation}
k_\lambda ^2-m^2+ie\left( \sigma \frac 1{F^{(\lambda )}}{\cal G}
-gF^{(\lambda )}\right) =eH(2n+1),\hspace{0.3in}n=1,2,...,  \label{33}
\end{equation}

$n$ is the principal quantum number and $k_\lambda $ is spectral parameter.
Equation (31) has four solutions with different asymptotics at $t\rightarrow
\pm \infty $ [34]
\[
_{+}\chi ^{(\lambda )}(\tau )=D_\nu [-(1-i)\tau ],\hspace{0.3in}^{-}\chi
^{(\lambda )}(\tau )=D_\nu [(1-i)\tau ],
\]
\vspace{-8mm}
\begin{equation}
\label{34}
\end{equation}
\vspace{-8mm}
\[
^{+}\chi ^{(\lambda )}(\tau )=D_{\nu ^{*}}[(1+i)\tau
],\hspace{0.3in} _{-}\chi ^{(\lambda )}(\tau )=D_{\nu
^{*}}[-(1+i)\tau ],
\]

where $D_\nu (x)$ are the parabolic-cylinder functions (the Weber-Hermite
functions) and
\[
\nu =\frac{ik_\lambda ^2}{2eE}-\frac 12.
\]

The four solutions of equation (25) for the potential (26) with different
asymptotic forms are given by

\begin{equation}
_{\pm }^{\pm }\psi _\mu ^{(\lambda )}(x)=N_0\xi _\mu \exp \left\{
i(p_2x_2+p_3x_3)-\frac{\eta ^2}2\right\} H_n(\eta )_{\pm }^{\pm }\chi
^{(\lambda )}(\tau ).  \label{35}
\end{equation}

Exact solutions (35) will be used to estimate the pair production
probability of vector particles and antiparticles in the
external constant and uniform electromagnetic fields.

\section{Pair production of vector particles with EDM and AMM}

The probability for pair production of vector particles with the EDM and AMM
by constant electromagnetic fields can be obtained through the asymptotic
form of solutions (35) when the time $t\rightarrow \pm \infty .$ The
functions $_{+}^{+}\psi ^{(\lambda )}(\tau )$ at $t\rightarrow \pm \infty $
have positive frequency and $_{-}^{-}\psi ^{(\lambda )}(\tau )$ have
negative frequency. Three quantities $k_\lambda ^2$ and the momentum
projections $p_2$, $p_3$ entering solutions (34), (35) are conserved. The
functions (34) (see [34]) obey the relations
\[
_{+}\chi ^{(\lambda )}(\tau )=c_{1n\lambda }{}^{+}\chi ^{(\lambda )}(\tau
)+c_{2n\lambda }{}^{-}\chi ^{(\lambda )}(\tau ),
\]
\[
^{+}\chi ^{(\lambda )}(\tau )=c_{1n\lambda }^{*}{}_{+}\chi ^{(\lambda
)}(\tau )-c_{2n\lambda }{}_{-}\chi ^{(\lambda )}(\tau ),
\]
\vspace{-8mm}
\begin{equation}
\label{36}
\end{equation}
\vspace{-8mm}
\[
^{-}\chi ^{(\lambda )}(\tau )=-c_{2n\lambda }^{*}{}_{+}\chi
^{(\lambda )}(\tau )+c_{1n\lambda }{}_{-}\chi ^{(\lambda )}(\tau
),
\]
\[
_{-}\chi ^{(\lambda )}(\tau )=c_{2n\lambda }^{*}{}^{+}\chi ^{(\lambda
)}(\tau )+c_{1n\lambda }^{*}{}^{-}\chi ^{(\lambda )}(\tau ),
\]

where
\[
c_{2n\lambda }=\exp \left[ -\frac \pi 2(\varepsilon +i)\right] ,
\hspace{0.3in}\varepsilon =\frac{m^2-ie\left( \sigma \frac 1{F^{(\lambda )}}
{\cal G}-gF^{(\lambda )}\right) +eH(2n+1)}{eE},
\]
\vspace{-8mm}
\begin{equation}
\label{37}
\end{equation}
\vspace{-8mm}
\[
\mid c_{1n\lambda }\mid ^2-\mid c_{2n\lambda }\mid ^2=1.
\]

The value $c_{2n\lambda }$ allows us to calculate the probability of pair
production of vector particles in the state with the quantum number $n$ and
corresponding to the eigenvalue $F^{(\lambda )}$. The probability for the
production of a pair of vector particles in the state with quantum number $n$
, components of momentum $p_2$, $p_3$ and corresponding to the eigenvalue $
F^{(\lambda )}$ throughout all space and during all time is

\begin{equation}
\mid c_{2n\lambda }\mid ^2=\exp \left\{ -\pi \left[ \frac{m^2}{eE}+\frac
HE(2n+1)\right] \right\} \mid \exp \left[ i\pi \left( \sigma \frac
1{F^{(\lambda )}}{\cal G}-gF^{(\lambda )}\right) /E\right] \mid .
\label{38}
\end{equation}

The expression (38) gives also the probability of the annihilation of a pair
of particles with quantum numbers $n,$ $p_2,$ $p_3$. From Eq. (38) we find
the average number of pairs produced from a vacuum

\begin{equation}
\overline{N}=\int \sum_{n,\lambda }\mid c_{2n\lambda }\mid ^2dp_2dp_3\frac{
L^2}{(2\pi )^2},  \label{39}
\end{equation}

where $(2\pi )^{-2}dp_2dp_3L^2$ means the final state density with the
cut-off $L$ along the coordinates ($V=L^3$ is the normalization volume). In
accordance with the approach [34] we can use the substitutions

\begin{equation}
\int dp_2\rightarrow eHL,\hspace{0.3in}\int dp_3\rightarrow eET.  \label{40}
\end{equation}

Here $T$ is the time of observation. It is possible to calculate the sum in
(39) over the principal quantum number $n$, and eigenvalues $\lambda $ with
the help of Eqs. (38), (23). Using Eqs. (40) we obtain the probability of
pair production of particles per unit volume and per unit time

\begin{equation}
I(E,H)=\frac{\overline{N}}{VT}=\frac{e^2EH}{8\pi ^2}\frac{\exp \left[ -\pi
m^2/(eE)\right] }{\sinh \left( \pi H/E\right) }\sum_\lambda \mid \exp \left[
i\pi \left( \sigma \frac 1{F^{(\lambda )}}{\cal G}-gF^{(\lambda )}\right)
/E\right] \mid .  \label{41}
\end{equation}

Evaluating the sum with the help of Eqs. (23)

\begin{equation}
\sum_\lambda \mid \exp \left[ \pi \left( \sigma \frac 1{F^{(\lambda )}}
{\cal G}-igF^{(\lambda )}\right) /E\right] \mid =2\cosh \pi \left( \sigma
+g\frac HE\right) +2,  \label{42}
\end{equation}

we arrive at the pair production probability

\begin{equation}
I(E,H)=\frac{e^2EH}{4\pi ^2}\frac{\cosh \pi \left( \sigma +gH/E\right) +1}{
\sinh \left( \pi H/E\right) }\exp \left[ -\pi m^2/(eE)\right] .  \label{43}
\end{equation}

So $I(E,H)$ is the intensity of the creation of pairs of particles with
spins of $1$, $0$. Below we extract the pair production probability for
particles with the pure spin $1$ possessing the gyromagnetic ratio $g$ (and
magnetic moment $\mu =eg/(2m)$) and the EDM $\sigma /(2m)$.

It follows from (43) that there is a pair production in a purely magnetic
field if $g>1$ that indicates about the instability of the vacuum in the
magnetic field. For the case $g>1$, $\sigma =0$ this property for the higher
spin particles was pointed in [16].

It is interesting to compare the probability (43) with those for particles
possessing pure spin $1$. The probability of pair production per unit volume
and per unit time of vector particles on the base of $\left( 0,1\right)
\oplus \left( 1,0\right) $ representation of the Lorentz group at $\sigma =0$
is given by [16, 17]
\begin{equation}
I^{(1)}(E,H)=\frac{e^2EH}{8\pi ^2}\frac{\exp \left[ -\pi m^2/(eE)\right] }{
\sinh \left( \pi H/E\right) }\frac{\sinh \left[ 3g\pi H/(2E)\right] }{\sinh
\left[ g\pi H/(2E)\right] }.  \label{44}
\end{equation}

Setting $\sigma =0$ in (43) and using some transformations we arrive at the
equality
\begin{equation}
I(E,H)=I^{(1)}(E,H)+I^{(0)}(E,H),  \label{45}
\end{equation}

where
\[
I^{(0)}(E,H)=\frac{e^2EH}{8\pi ^2}\frac{\exp \left[ -\pi m^2/(eE)\right] }{
\sinh \left( \pi H/E\right) }
\]
is the intensity of the creation of pairs of scalar particles [13] (see also
the creation of pairs of composite scalar particles in [36]). The physical
meaning of equation (45) is clear: the probability of pair production of
fields with spins $1$, $0$ is the sum of production probabilities of vector
and scalar particles. By excepting Eq. (45) for arbitrary $\sigma $ and $g$
we obtain from Eq. (43) the expression for pair production probability of
particles with pure spin one:
\begin{equation}
I^{(1)}(E,H)=\frac{e^2EH}{8\pi ^2}\frac{2\cosh \pi \left( \sigma
+gH/E\right) +1}{\sinh \left( \pi H/E\right) }\exp \left[ -\pi
m^2/(eE)\right] .  \label{46}
\end{equation}

The Eq. (46) obtained is a new result for the intensity of pair production
of vector particles with the EDM and AMM. From general formula (46) we find
that in the case $\sigma =g=0$, the pair production of vector particles is
three times that for scalar pair production. This is due to the three
physical degrees of freedom of the vector field.

The imaginary part of the density of the Lagrangian can be obtained using
the relationship [34]:

\begin{equation}
VT \mbox{Im} {\cal L}=\frac 12\int \sum_{n,\lambda }\ln \mid
c_{1n\lambda }\mid ^2dp_2dp_3\frac{L^2}{(2\pi )^2}.  \label{47}
\end{equation}

From Eq. (47) taking into account Eqs. (37), (40) we arrive at

\begin{equation}
\mbox{Im} {\cal L}=\frac{e^2EH}{8\pi ^2}\sum_{k=1}^\infty
\frac{(-1)^{k-1}} k\exp \left( -\frac{\pi km^2}{eE}\right)
\frac{\cosh \pi k\left( \sigma +gH/E\right) +1}{\sinh \left( \pi
kH/E\right) }.  \label{48}
\end{equation}

According to the approach [13] the first term in (48) (at $k=1$) coincides
with the intensity of pair production (43) (probability of the pair production
per unit volume per unit time) divided by 2. Expression Im$\mathcal{L}$ (48)
and the pair production probability (46) do not depend on the
renormalization scheme because all divergences and the renormalizability are
contained in Re$\mathcal{L}$ [13].

\section{Polarization of vector particle vacuum}

In this section we evaluate one-loop corrections to the Lagrange function of
a constant and uniform electromagnetic field due to the field interaction
with a vacuum of vector particles with the EDM and AMM. This problem has
been solved for a number of theories [13, 15-18]. The effect of scattering
of light by light is described by the nonlinear corrections to the
Lagrangian of the electromagnetic field. Adapting the Schwinger method [13]
to the fields described by equation (22), we obtain the nonlinear
corrections to Lagrangian of a constant and uniform electromagnetic field

\begin{equation}
{\cal L}^{(1)}=\frac 1{16\pi ^2}\int_0^\infty d\tau \tau ^{-3}\exp
\left( -m^2\tau -l(\tau )\right) \mbox{tr}\exp \left[ ie_0\left(
\sigma \widetilde{F }_{\mu \nu }-gF_{\mu \nu }\right) \tau \right]
,  \label{49}
\end{equation}

with

\begin{equation}
l(\tau )=\frac 12\mbox{tr}\ln \left[ \left( e_0F\tau \right)
^{-1}\sin (e_0F\tau )\right] ,\hspace{0.3in}\exp \left[ -l(\tau
)\right] =\frac{ (e_0\tau )^2{\cal G}_0}{\mbox{Im}\cosh (e_0\tau
X_0)}, \label{50}
\end{equation}

where ${\bf X}_0={\bf H}_0+i{\bf E}_0$, $X=\sqrt{{\bf X}^2}$, $
{\cal G}_0={\bf E}_0{\bf H}_0$, ${\bf E}_0$, ${\bf H}_0$ are
bare (nonrenormalized) electric and magnetic fields, respectively, $e_0$ is
the bare electric charge (the index $0$ refers to the unrenormalized
variables)$.$ The expression (49) is the effective nonlinear Lagrangian
which is represented as an integral over the proper time $\tau $. Here we
consider general case of arbitrary constant vectors ${\bf E}_0$ and $
{\bf H}_0$. With the help of Eqs. (23) we calculate the trace (tr) of the
matrices
\[
\mbox{tr}\exp \left[ ie_0\left( \sigma \widetilde{F}_{\mu \nu
}-gF_{\mu \nu }\right) \tau \right]
\]
\vspace{-8mm}
\begin{equation}
\label{51}
\end{equation}
\vspace{-8mm}
\[
=2\left[ \cosh e_0\tau \left( \frac{\sigma {\cal G}_0}{\mbox{Re}
X_0}+g \mbox{Re} X_0\right) +\cos e_0\tau \left( \frac{\sigma
{\cal G}_0}{\mbox{Im} X_0}-g \mbox{Im} X_0\right) \right] .
\]

Substituting (51) into (49) and subtracting the additive constant to ensure
that the expression ${\cal L}^{(1)}$ vanishes for zero fields, we get
\[
{\cal L}^{(1)}=\frac 1{8\pi ^2}\int_0^\infty d\tau \tau ^{-3}\exp \left(
-m^2\tau \right) \times
\]
\vspace{-8mm}
\begin{equation}
\label{52}
\end{equation}
\vspace{-8mm}
\[
\times \left[ (e_0\tau )^2{\cal G}_0\frac{\cosh e_0\tau \left(
\sigma {\cal G}_0/ \mbox{Re} X_0+g \mbox{Re} X_0\right) +\cos
e_0\tau \left( \sigma {\cal G}_0/ \mbox{Im} X_0-g \mbox{Im}
X_0\right) }{ \mbox{Im} \cosh (e_0\tau X_0)}-2\right] ,
\]

The integral (52) is the nonlinear correction to Maxwell's
Lagrangian due to the vacuum polarization of vector (with the
additional scalar field) fields which possess the EDM and AMM. The
Lagrangian (52) contains the term that renormalizes the Lagrangian
of the free electromagnetic fields
\begin{equation}
{\cal L}^{(0)}=-{\cal F}_0=\frac 12\left( {\bf E}_0^2- {\bf
H}_0^2\right) .  \label{53}
\end{equation}

Extracting the divergent constant in Eq. (52) for weak fields, and adding
Eq. (52) to the Maxwell Lagrangian (53) we obtain the renormalized
Lagrangian of electromagnetic fields
\[
{\cal L}={\cal L}^{(0)}+{\cal L}^{(1)}=-{\cal F}+\frac 1{8\pi
^2}\int_0^\infty d\tau \tau ^{-3}\exp \left( -m^2\tau \right) \times
\]
\begin{equation}
\times \biggl [ (e\tau )^2{\cal G}\frac{\cosh e\tau \left( \sigma
{\cal G}/\mbox{Re} X+g \mbox{Re} X\right) +\cos e\tau \left(
\sigma {\cal G}/ \mbox{Im} X-g \mbox{Im} X\right) }{ \mbox{Im}
\cosh (e\tau X)}
\end{equation}\label{54}
\[
-2+(e\tau )^2\left( \frac 23+\sigma ^2-g^2\right) {\cal F}\biggr ] ,
\]

where the renormalized fields and charges are used:
\[
{\cal F}=Z_3^{-1}{\cal F}_0,\hspace{0.3in}e=Z_3^{1/2}e_0,
\]

and the renormalization constant is given by

\begin{equation}
Z_3^{-1}=1+\frac{e_0^2}{12\pi ^2}\left[ 1+\frac 32\left( \sigma
^2-g^2\right) \right] \int_0^\infty d\tau \tau ^{-1}\exp \left( -m^2\tau
\right) .  \label{55}
\end{equation}

The integral (54) vanishes already if the electromagnetic fields ${\bf E}$
, ${\bf H}$ are absent. We can use the cutoff factor $\tau _0$ at the
lower limit in the integral (55), and the constant $Z_3^{-1}$ diverges
logarithmically as $\tau _0\rightarrow 0$. When the EDM is absent ($\sigma
=0 $) and the gyromagnetic ratio $g=2$ that is the linear approximation to
the renormalizable gauge theory, we arrive from Eq. (55) to the
renormalization constant obtained in [15]. It follows from Eq. (55) that
when the inequality
\begin{equation}
g^2-\sigma ^2>\frac 23  \label{56}
\end{equation}

is valid the renormalization constant of the charge $Z_3^{1/2}$
becomes larger than one. This case, unlike ordinary
electrodynamics, corresponds to the absence of the zero charge
situation in the asymptotic region and indicates asymptotic
freedom in the field [37, 38]. According to Eq. (56) the
asymptotically free behavior in the vector field is due to the AMM
but the role of the EDM is opposite. At the case $\sigma
^2-g^2>2/3$ the situation of the zero charge situation in the
asymptotic region, like electrodynamics, is realized. From Eq.
(55) we obtain the Callan-Zymanzik $ \beta$-function that
corresponds to the renormalizable theory
\begin{equation}
\beta =\frac{e_0^2}{12\pi ^2}\left[ 1+\frac 32\left( \sigma ^2-g^2\right)
\right] .  \label{57}
\end{equation}

At the condition (56) the $\beta $-function is negative ($\beta
<0$) and we arrive at the region of asymptotic freedom. The AMM
assures asymptotic freedom and instability of the vacuum in a
magnetic field.

Expanding Eq. (54) in the small electromagnetic fields we obtain the Maxwell
Lagrangian including the nonlinear corrections (in rational units)
\[
{\cal L}=\frac 12\left( {\bf E}^2-{\bf H}^2\right) +\frac{6
\sigma g}{2+3(\sigma ^2-g^2)}\left( {\cal G}-{\cal G}_0\right)
\]
\begin{equation}
+\frac{\alpha ^2}{m^4}\biggl [
\frac{14-30(g^2-\sigma ^2)+15(g^4+\sigma ^4)}{45}{\cal F}^2
\end{equation}\label{58}
\[
+\frac 2{45}\left( 1+\frac{15(\sigma ^4+6\sigma ^2g^2+g^4)}4\right) {\cal
G}^2+\frac 23\sigma g\left( g^2-\sigma ^2-2\right) {\cal GF}\biggr ] .
\]
where $\alpha =e^2 /(4\pi)$. The second term in Eq. (58) is
induced parity violation anomaly for a vector field with the EDM.
This and last terms in Eq. (58) violate parity symmetry due to the
EDM of a particle. The effective Lagrangian (58) is like the
Heisenberg-Euler Lagrangian of QED [39, 40] but in the case of the
polarized vacuum of vector fields with arbitrary EDM and AMM and
additional scalar field (with the same mass). The presence of a
scalar field is due to the special gauge $\xi =1$ which was chosen
to simplify the calculations. Now we will take into consideration
the contribution of a scalar (nonphysical) field. It is easy to
verify that for the particular case of $\sigma =0,$ $ g=0 $,
Eq.(58) becomes
\[
{\cal L}(\sigma =g=0)=\frac 12\left( {\bf E}^2-{\bf H}^2\right) +
\frac{\alpha ^2}{90m^4}\left[ 7\left( {\bf E}^2-{\bf H}^2\right) ^2+4(
{\bf EH})^2\right]
\]
\vspace{-8mm}
\begin{equation}
\label{59}
\end{equation}
\vspace{-8mm}
\[
=\frac 12\left( {\bf E}^2-{\bf H}^2\right) +4{\cal L}_{spin 0},
\]
where
\begin{equation}
{\cal L}_{spin 0}=\frac{\alpha ^2}{360m^4}\left[ 7\left( {\bf E}
^2-{\bf H}^2\right) ^2+4({\bf EH})^2\right]  \label{60}
\end{equation}
is the correction to the Maxwell Lagrangian due to the vacuum
polarization of scalar pointlike particles [13]. As Eq. (22)
becomes a Klein-Gordon equation for the field $\psi _\mu $ at
$\sigma =g=0$, there is an equal contribution of four degrees of
freedom of fields with spins $1$ (three projection $\pm 1$, $0$)
and $0$. To have the contribution from a field of pure spin $1$ we
should subtract from Eq. (58) the expression (60) corresponding to
spin $0$ of a field. As a result the Lagrangian of a constant,
uniform, electromagnetic field taking into account the vacuum
polarization of a charged vector particles with the EDM and AMM is
given by
\[
{\cal L}_{spin 1}={\cal L}-{\cal L}_{spin 0}=\frac
12\left( {\bf E}^2-{\bf H}^2\right) +\frac{6\sigma g}{2+3(\sigma
^2-g^2)}\left( {\bf EH-E}_0{\bf H}_0\right)
\]
\begin{equation}
+\frac{\alpha ^2}{m^4}\biggl [
\frac{7-20(g^2-\sigma ^2)+10(g^4+\sigma ^4)}{120}\left( {\bf E}^2-{\bf
H}^2\right) ^2
\end{equation}\label{61}
\[
+\frac{1+5(\sigma ^4+6\sigma ^2g^2+g^4)}{30}({\bf EH})^2+\frac
13\sigma g\left( g^2-\sigma ^2-2\right) ({\bf EH})\left( {\bf
E}^2-{\bf H} ^2\right) \biggr ] .
\]

For the particular case $\sigma =0,$ $g=2$ which corresponds to the linear
approximation to the renormalizable SM Eq. (61) leads to the expression
\begin{equation}
{\cal L}_{spin 1}=\frac 12\left( {\bf E}^2-{\bf H}^2\right) +\frac{\alpha ^2
}{10m^4}\left[ \frac{29}4\left( {\bf E}^2-{\bf H}^2\right) ^2+27(
{\bf EH})^2\right]  \label{62}
\end{equation}

which coincides with those obtained in [15].

It is possible to obtain the asymptotic form of (54) for super-critical
fields at $eE/m^2\rightarrow \infty $ and $eH/m^2\rightarrow \infty $.
However, for strong electromagnetic fields the AMM and EDM can depend on the
external field like the dependence of the electron AMM in QED [41, 42].

\section{Conclusion}

Starting with the Dirac-K\"ahler equation for tensor fields we
arrived at the two $P$-odd subsystem for self-dual and
antiselfdual antisymmetric tensors of second rank. These equations
are based on the $(0,0)\oplus (1/2,1/2)\oplus $ $\left( 1,0\right)
$ and $(0,1)\oplus (1/2,1/2)\oplus (0,0) $ representations of the
Lorentz group and describe fields with spins of $1$ and $0$. The
8-component Dirac-like $P$-odd matrix wave equation is constructed
possessing $GL(2,c)$ group of symmetry. This symmetry is due to
the presence of two spins $1$ and $0$. The system of tensor
equations considered allows us to introduce the EDM and AMM of a
particle in the first order equations. The second order equation
for a particle with the EDM and AMM is simpler (for solving)
compared to the Proca equation. This equation can be treated as an
equation for a vector particle with the gauge $\xi =1$ in the
framework of the T. D. Lee and C. N. Yang formalism. The
contribution of the nonphysical scalar field to physical
observables is eliminated at the end of calculations. Such
approach allowed us to obtain the pair-production probability, and
the effective Lagrangian for electromagnetic fields taking into
account the polarization of the vacuum of vector particles with
the EDM and AMM. This is the generalization of the Schwinger
result on the case of vector particles in the external electric
and magnetic fields. The exact formula for the intensity of pair
production of fields with spins $1$ and $0$ is the sum of the
intensity of pair production of vector and scalar particles. It is
shown that there is a pair production of vector particles by a
purely magnetic field in the case of $g>1$ assuring asymptotic
freedom and instability of the vacuum in a magnetic field. The
role of the EDM of a vector particle is opposite: the EDM of a
particle does not lead to instability of the vacuum in a magnetic
field and suppresses the phenomena of asymptotic freedom. The pair
production probability does not depend on the renormalization
scheme because all divergences and the renormalizability are
contained in Re${\cal L}$. Discussing the procedure of the
renormalization we imply that the scheme considered is the
linearized version of renormalize gauge theory. This point of view
is due to the smallness of the vector field self-interaction
constant (see [14]) and it is possible to ignore processes that
allow for the self-interaction of the vector field in vacuum.

The presence of the EDM and the value of the AMM $\kappa \neq 1$ ($g\neq 2$)
of a vector particle lieds to physics beyond the SM. Recent experimental
muon AMM data [43] have challenged the SM as there is a discrepancy of $
2.6\sigma $ deviation between the theory and the averaged experimental
value. This can open a window to new physics.

\end{document}